      \documentclass[a4paper,11pt]{article}  

\usepackage{jheppub1}

\usepackage{float}
\usepackage{epstopdf}

\usepackage{hyperref} 
\usepackage{cleveref}

\title{\boldmath Greybody factor of scalar field from
Reissner-Nordstr{\"o}m-de Sitter black hole}

\author[a]{Jamil Ahmed} 
\author[a,b]{and K. Saifullah}

\affiliation [a]{Department of Mathematics, Quaid-i-Azam University, Islamabad, Pakistan}
\affiliation[b]{Center for the Fundamental Laws of Nature, Harvard University, Cambridge, MA 02138, USA \\ }

 \affiliation{\emph{Electronic Address: jahmed@student.qau.edu.pk; ksaifullah@fas.harvard.edu}}

\abstract{In this work we derive a general expression for the greybody factor of
non-minimally coupled scalar fields in Reissner-Nordstr{\"o}m-de Sitter
spacetime in low frequency approximation. Greybody factor as a characteristic of effective potential barrier, will be presented. We discuss the role of cosmological constant both, in the absence as well as in the presence of non-minimal coupling. Considering non-minimal coupling as a mass term, its effect on the greybody factor will be discussed. We also elaborate
the significance of the results by giving formulae of differential energy
rate and general absorption cross section. The greybody factor gives
insight into the spectrum of Hawking radiations.}

\begin{document}
\maketitle
\flushbottom

\section{Introduction}
\label{sec:intro}

The study of asymptotically non-flat spacetime geometries received a lot of
attention after it was discovered that our Universe has entered into a new
phase of accelerated expansion \cite{S}. Among these non-flat geometries de
Sitter spacetime is of great interest due to its rich symmetries and also
because it could incorporate the accelerated expansion of the Universe due to
the presence of non-zero cosmological constant in the Einstein field
equations. As predicted, the Universe is in continous expansion, so in far
future it will pass through a de Sitter phase. Further, de Sitter geometry
could also approximate the inflationary phase of our Universe \cite{A}. The
fact that all black hole spacetimes of Kerr-Newman family can be generalized
to include a cosmological constant makes black hole de Sitter spacetimes
an interesting field of investigation. De Sitter spacetime is the maximally
symmetric Lorentzian space having positive curvature. In four dimensions the
symmetry group is $SO(1,4)$ and topology is $R\times S$. Due to the
structure of de Sitter spacetime inertial observers are surrounded by
cosmological horizons, which are a characteristic of spacetimes having positive
cosmological constant. Also dS/CFT correspondence enhances the interest in
the study of de Sitter spacetimes as they provide connection with conformal
field theories. The absorption and emission spectra of a Schwarzschild black hole was studied in Ref. \cite{NS}. There has been a considerable interest in the study of
greybody factor of scalar and fermionic radiations from asymptotically flat
spacetimes and black strings \cite{C, SD, LSE, YH, SJ, M,JA,SI,IS,MF,WJ,ahmed2017greybody}.

Black hole emission and absorption phenomena is related to this important
quantity known as greybody factor. It is this quantity that makes it
different from emission and absorption of black body. The question is, how this
quantity originated? It is generated by an effective potential barrier by
black hole spacetimes. This potential quantum mechanically allows some of the
radiation to transmit and remaining to reflect back. This leads to the
frequency dependent greybody factor. Due to this factor black hole thermal
radiation formula is different from the black body radiation formula.
Greybody factor not only alters the thermal radiation formula but is also
important to compute the partial absorption cross section of black holes
\cite{B, D, W}. In the literature there are investigations for greybody factor
of scalar fields for Schwarzschild-de Sitter black hole. These include the
cases of lowest partial modes in low energy regimes \cite{PJA, TJR, LAE, PTN}. 

In this paper we use the simple matching technique to solve the radial
equation resulting from the Klein Gordon equation in the background of the
Riessner-Nordstr{\"o}m-de Sitter black hole. In this method we divide the space
into two regimes, namely near the black hole horizon and near the
cosmological horizon and find solutions for radial equations in both the
regimes separately. Then we stretch these solutions to an intermediate point
$r_{m}$. The choice of $r_{m}$ is such that
\begin{equation}
r_{h}<r_{m},\text{ \ \ }r_{c}>r_{m}\text{ \ \ and }\omega r_{m}\ll 1,
\label{1a}
\end{equation}%
where $r_{h}$ and $r_{c}$ correspond to black hole and cosmological
horizons respectively and $\omega $ is the frequency.

The rest of the paper is organized as follows. In Section \ref{sec:KGEffectPot} we will
discuss the Klein Gordon equation and the profile of effective potential in the
background of Reissner-Nordstr{\"o}m-de Sitter black hole. In Section \ref{sec:GBFCom} we
compute the greybody factor, starting from near black hole horizon solution,
near cosmological horizon solution, and then matching them at an intermediate point.
This yields expressions for greybody factor and spectrum of Hawking
radiations. At the end we make some concluding remarks.

\section{Klein Gordon equation and profile of effective potential} \label{sec:KGEffectPot}

The spacetime metric for Reissner-Nordstr{\"o}m-de Sitter black hole is given
by,
\begin{equation}
ds^{2}=g(r)dt^{2}-\frac{1}{g(r)}dr^{2}-r^{2}\left( d\theta ^{2}+\sin
^{2}\theta d\varphi ^{2}\right) ,  \label{1}
\end{equation}
where 
\begin{equation}
g(r)=1-\frac{2 M}{r}+\frac{Q^{2}}{r^{2}}-\frac{\Lambda r^{2}}{3}.  \label{2}
\end{equation}
and related electromagnetic field is given by the four-potential
\begin{equation}
A_{\mu }=\frac{Q}{r}\delta _{\mu }^{t}.  \label{1aa}
\end{equation}%
Here $M$ is mass and $Q$ is charge of the black hole. Introducing a
dimensionless cosmological parameter $\lambda =\frac{1}{3}\Lambda M^{2}$, a
dimensionless charge $e=\frac{Q}{M}$ and dimensionless coordinates $%
t\longrightarrow t/M$, $r\longrightarrow r/M$. It is equivalent to putting $M=1$ 
\cite{ZS11}. The horizons are determined by the condition
\begin{equation}
1-\frac{2}{r}+\frac{e^{2}}{r^{2}}-\lambda r^{2}=0.  \label{1ab}
\end{equation}%
It is clear from above equation that in the special case of $e=0$, the black
hole spacetimes exist for all $\lambda \leqslant 0$ and for $0<\lambda
\leqslant \frac{1}{27}$.

Before going into detailed calculations of analytic result of greybody factor
we comment about its validity. It is interesting to note that it is valid for
arbitrary quantum number $l$ and coupling $\xi $. On the other hand the
accuracy of the result is guaranteed only if the two asymptotic regions
overlap, which implies that it is only valid for small frequencies. Also the
approximation which we have used is justified for ``small'' black holes
(compared with characteristic dS scale) that is $\lambda <<1$. Therefore the
result of greybody factor is valid only in complementary regions of
parameter space.

We consider a scalar field theory in which the field is either minimally or non-minimally coupled to gravity and described by the following action 
\begin{equation}\label{actionscalar}
\mathcal{S} =\int{d^{4}x \sqrt{-g}[R-\xi R \Phi^{2}-\partial_{\mu}\Phi \partial^{\mu}\Phi].}
\end{equation}
The equation of motion for the above theory can be written as
\begin{equation}
\frac{1}{\sqrt{-g}}\partial _{\mu }\left[ \sqrt{-g}g^{\mu \nu }\partial
_{\nu }\Phi (t,r,\theta ,\varphi )\right] =-4 \xi \Lambda \Phi .  \label{3}
\end{equation}
In above, we have used $R=-4 \Lambda$ and $\xi $ is a coupling constant determining the magnitude of coupling between the scalar and gravitational field, with $\xi=0$ corresponding to the minimal coupling. In matrix form the above line element can be written as
\begin{equation}
g_{\mu \nu }=\left(
\begin{array}{cccc}
g(r) & 0 & 0 & 0 \\
0 & -\frac{1}{g(r)} & 0 & 0 \\
0 & 0 & -r^{2} & 0 \\
0 & 0 & 0 & -r^{2}\sin ^{2}\theta
\end{array}%
\right) .  \label{4}
\end{equation}
Also,
\begin{equation}
\sqrt{-g}=r^{2}\sin \theta .  \label{5}
\end{equation}
Using these values in equation $\left(\ref{3}\right) $, it takes the form
\begin{equation*}
\frac{1}{r^{2}\sin \theta }\partial _{t}\left( \frac{r^{2}\sin \theta }{g(r)}%
\partial _{t}\Phi \right) +\frac{1}{r^{2}\sin \theta }\partial _{r}\left(
r^{2}\sin \theta \left( -g(r)\right) \partial _{r}\Phi \right) +\frac{1}{%
r^{2}\sin \theta }\partial _{\theta }\left( -\sin \theta \partial _{\theta
}\Phi \right)
\end{equation*}%
\begin{equation}
+\frac{1}{r^{2}\sin \theta }\partial _{\varphi }\left( \partial _{\varphi
}\Phi \right) =-4 \xi \Lambda \Phi .  \label{7}
\end{equation}
Let%
\begin{equation}
\Phi (t,r,\theta ,\varphi )=e^{-\iota \omega t}R(r)Y(\theta ,\varphi ),
\label{8}
\end{equation}
therefore, the radial part of equation $\left( \ref{7}\right) $ is 
\begin{equation}
\frac{1}{r^{2}}\frac{d}{dr}\left( r^{2}g(r)\right) \frac{dR(r)}{dr}+\left[
\frac{\omega ^{2}}{g(r)}-\frac{l(l+1)}{r^{2}}+4 \xi \Lambda \right] R(r)=0,
\label{9}
\end{equation}
where $l(l+1)$ are the eigenvalues coming from the $(\theta ,\varphi )$ part.

Before solving equation $\left( \ref{9}\right) $ we will discuss the profile
of effective potential due to which greybody factor originates. We employ
the following transformation on equation $\left( \ref{9}\right) $
\begin{equation}
R\left( r\right) =\frac{U\left( r\right) }{r}  \label{10}
\end{equation}
and the tortoise coordinate
\begin{equation}
x\equiv \int \frac{dr}{g},  \label{11}
\end{equation}
such that
\begin{equation*}
\frac{d}{dx}=g\frac{d}{dr},\frac{d^{2}}{dx^{2}}=g^{2}\frac{d^{2}}{dr^{2}}%
+gg^{\prime }\frac{d}{dr}.
\end{equation*}
Thus equation $\left( \ref{9}\right) $ takes the form
\begin{equation}
\left( \frac{d^{2}}{dx^{2}}+\omega ^{2}-V_{eff}\left( r\right) \right)
U\left( r\right) =0,  \label{12}
\end{equation}
with
\begin{equation}
V_{eff}\left( r\right) =g\left( r\right) \left( \frac{l\left( l+1\right) }{%
r^{2}}-4 \xi \Lambda+\frac{g^{\prime }}{r}\right) .  \label{13}
\end{equation}
\begin{figure}[!h]
  \centering
  \includegraphics[width=0.6\textwidth]{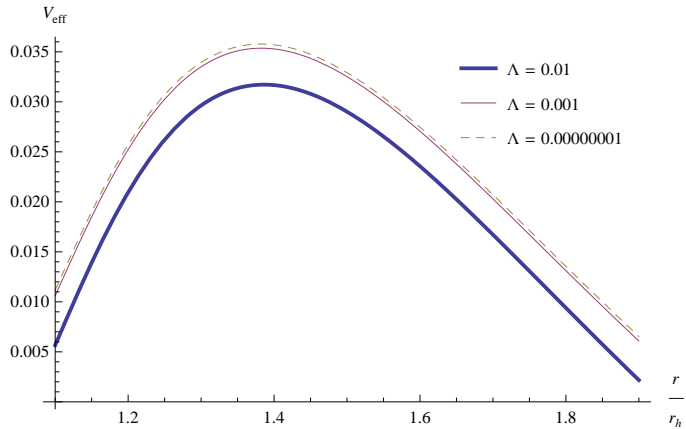}
  \caption{Profile of effective potential for different values of the cosmological constant for $\xi=0.01$, $q=1$ and $l=0$}
  \label{Effective-potential}
  \end{figure}
\begin{figure}[!h]
  \centering
  \includegraphics[width=0.6\textwidth]{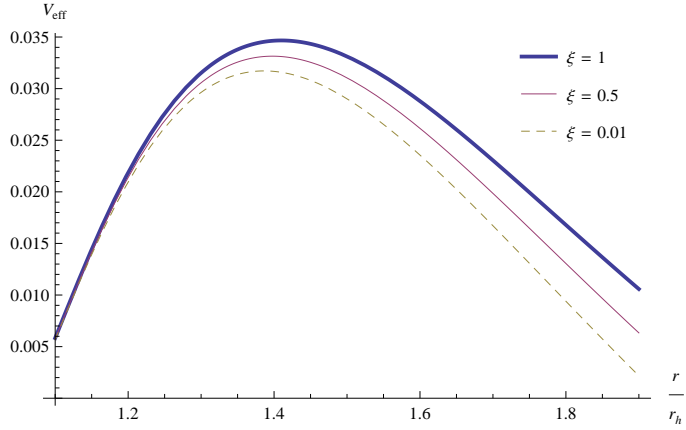}
  \caption{Profile of effective potential for different values of the coupling constant for $\Lambda=0.01$, $q=1$ and $l=0$}\label{Effective-potential-1}
  \end{figure}

In Fig. \ref{Effective-potential}, we draw the profile of effective potential for different values of the cosmological constant for $\xi=0.01$, $q=1$ and $l=0$. It is observed that an increase in the value of cosmological constant, decreases the height of gravitational barrier and thus enhances the greybody factor. In Fig. \ref{Effective-potential-1}, the effective potential is depicted for different values of the coupling constant and $\Lambda=0.01$, $q=1$ and $l=0$. It is observed that increase in the value of the coupling constant leads to the increase in gravitational barrier, which subsequently suppresses the emission of scalar fields.
\section{Greybody factor computation} \label{sec:GBFCom}

\subsection{Near black hole horizon solution}

Equation\ $\left( \ref{9}\right) $ is the master equation of our interest.
We will solve this equation in two regions separately, namely, near the black hole
horizon and the cosmological horizon by using a semi-classical approach known as
simple matching technique. Then we will match both the solutions to an
intermediate region to get the analytical expression for the greybody factor.

For the near horizon region $r\sim r_{h}$, we will perform the following
transformation to simplify the radial equation \cite{PTN}
\begin{equation}
r\rightarrow f=\frac{g\left( r\right) }{1-\Xi r^{2}},  \label{14}
\end{equation}
where
\begin{equation*}
\Xi =\frac{\Lambda }{3}.
\end{equation*}
Thus we get
\begin{equation}
\frac{df}{dr}=\left( 1-f\right) \frac{B(r_{h)}}{r_{h}\left( 1-\Xi
r_{h}^{2}\right) },  \label{15}
\end{equation}
where, in the above 
\begin{equation}
B(r_{h)}=\left(\frac{-6 \Lambda M r_{h}^{3}+4 \Lambda Q^{2} r_{h}^{2}+6 M r_{h}-6 Q^{2}}{6 M r_{h}-3 Q^{2}}\right).  \label{16}
\end{equation}
Using equations $\left( \ref{14}\right) $ and $\left( \ref{15}\right) $ in
$\left( \ref{9}\right) $, we obtain
\begin{equation}
f(1-f)\frac{d^{2}R(f)}{df^{2}}+\left( 1-C_{\ast }f\right) \frac{dR(f)}{df}+%
\left[ \frac{F_{\ast }^{2}}{B^{2}(r_{h)}\left( 1-f\right) f}-\frac{\lambda
_{h}\left( 1-\Xi r^{2}_{h}\right) }{B^{2}(r_{h)}\left( 1-f\right) }\right]
R(f)=0.  \label{17}
\end{equation}
Here
\begin{equation}
F_{\ast }=\omega r_{h},  \label{18}
\end{equation}%
\begin{equation}
C_{\ast }=\frac{r_{h}\left( 1-\Xi r^{2}_{h}\right) }{\left( 1-f\right)
^{2}B\left( r_{h}\right) }-\frac{2r_{h}^{2}\Xi \left( 1-\Xi r^{2}_{h}\right) }{%
\left( 1-f\right) B\left( r_{h}\right) },  \label{19}
\end{equation}
and
\begin{equation}
\lambda _{h}=\left[ l\left( l+1\right) -4 \xi \Lambda\right]
r_{h}^{2}.  \label{20}
\end{equation}
In order to further simplify the above equation we use field redefinition
\begin{equation}
R(f)=f^{\mu _{1}}\left( 1-f\right) ^{\nu _{1}}F(f).  \label{21}
\end{equation}
In equation $\left( \ref{17}\right) $ we use this definition of $R(f)$ to get

\begin{equation*}
f\left( 1-f\right) \frac{d^{2}F(f)}{df^{2}}+\left[ 1+2\mu _{1}-\left( 2\mu
_{1}+2\nu _{1}+C_{\ast }\right) f\right] \frac{dF}{df}+(\frac{\mu _{1}^{2}}{f%
}-\mu _{1}^{2}+\mu _{1}-2\mu _{1}\nu _{1}+\frac{\nu _{1}^{2}}{1-f}-\nu
_{1}^{2}-\frac{2\nu _{1}}{1-f}
\end{equation*}%
\begin{equation}
+\nu _{1}-\mu _{1}C_{\ast }+\frac{\nu _{1}C_{\ast }}{1-f}-\nu _{1}C_{\ast }+%
\frac{F_{\ast }^{2}}{B^{2}(r_{h})f}+\frac{F_{\ast }^{2}}{B^{2}(r_{h})\left(
1-f\right) }-\frac{\lambda _{h}\left( 1-\Xi r^{2}_{h}\right) }{%
B^{2}(r_{h)}\left( 1-f\right) })F(f)=0  \label{22}
\end{equation}
Now, define
\begin{equation}
a_{1}=\mu _{1}+\nu _{1}+C_{\ast }-1,  \label{23}
\end{equation}%
\begin{equation}
b_{1}=\mu _{1}+\nu _{1},  \label{23a}
\end{equation}%
\begin{equation}
c_{1}=1+2\mu _{1}.  \label{23b}
\end{equation}
Also constraints coming from the coefficients of $F(f)$ give
\begin{equation}
\mu _{1}^{2}+\frac{F_{\ast }^{2}}{B^{2}(r_{h})}=0,  \label{24}
\end{equation}
and%
\begin{equation}
\nu _{1}^{2}+\nu _{1}(C_{\ast }-2)+\frac{F_{\ast }^{2}}{B^{2}(r_{h})}-\frac{%
\lambda _{h}\left( 1-\Xi r^{2}_{h}\right) }{B^{2}(r_{h)}}=0.  \label{25}
\end{equation}
From here we find the values of $\mu_{1} $ and $\nu_{1} $ as%
\begin{equation}
\mu _{1}=\pm \iota \frac{F_{\ast }}{B(r_{h})},  \label{26}
\end{equation}
and%
\begin{equation}
\nu _{1}=\frac{1}{2}\left[ \left( 2-C_{\ast }\right) \pm \sqrt{\left(
2-C_{\ast }\right) ^{2}-4\left( \frac{F_{\ast }^{2}}{B^{2}(r_{h})}-\frac{%
4\lambda _{h}\left( 1-\Xi r^{2}_{h}\right) \lambda_{h}}{B^{2}(r_{h)}}\right) }\right]
.  \label{27}
\end{equation}
Thus equation $\left( \ref{22}\right) $ by virtue of equations $\left( \ref%
{23}, \ref{23a}, \ref{23b}\right) $ and constraints $\left( \ref{24}, \ref{25}\right)$ becomes
\begin{equation}
f\left( 1-f\right) \frac{d^{2}F(f)}{df^{2}}+\left[ c_{1}-\left(
1+a_{1}+b_{1}\right) f\right] \frac{dF(f)}{df}-a_{1}b_{1}F(f)=0.  \label{28}
\end{equation}
In the near horizon region the solution can be written in the form of
general hypergeometric function, which has the form
\begin{equation}
R(f)_{NH}=A_{1}f^{\mu _{1}}(1-f)^{\nu _{1}}F\left( a,b,c;f\right)
+A_{2}f^{-\mu _{1}}(1-f)^{\nu _{1}}F\left( a-c+1,b-c+1,2-c;f\right)
\label{29}
\end{equation}%
where $A_{1\text{ }}$and $A_{2}$ are arbitrary constants. For the near horizon
case there exists no outgoing mode, we choose $A_{2}=0$, thus
we get

\begin{equation}
R(f)_{NH}=A_{1}f^{\mu _{1}}(1-f)^{\nu _{1}}F\left( a,b,c;f\right) .
\label{30}
\end{equation}

\subsection{Near cosmological horizon solution}

We now solve the radial equation $\left( \ref{9}\right) $ close to the
cosmological horizon $r_{c}$. In this case we choose the radial function $h\left( r\right)$, in place of $g\left( r\right)$, which is defined as
\begin{equation}
h\left( r\right) =1-\Xi r^{2}.  \label{31}
\end{equation}%
We employ the following transformation on equation $\left( \ref{9}\right) $
\begin{equation}
r\rightarrow h\left( r\right) ,  \label{32}
\end{equation}%
so that

\bigskip
\begin{equation}
\frac{dh}{dr}=\frac{\left( 1-h\right) }{r}\left(-2\right) .
\label{32a}
\end{equation}%
Using this, equation $\left( \ref{9}\right) $ becomes%
\begin{equation}
h(1-h)\frac{d^{2}R(h)}{dh^{2}}+\left( 1-2h\right) \frac{dR(h)}{dh}+\left[
\frac{F_{c}^{2}}{B_{c}^{2}\left( 1-h\right) h}-\frac{\lambda _{c}}{\left(
1-h\right) B_{c}^{2}}\right] R(h)=0,  \label{33a}
\end{equation}%
where, $F_{c}= \omega  r_{c}^2$, $B_{c}= -2$ and $\lambda_{c}=\left[l \left(l+1\right)-4 \xi \Lambda\right] r_{c}^2$.
We redefine the radial function as
\begin{equation}
R(h)=h^{\mu _{2}}\left( 1-h\right) ^{\nu _{2}}X\left( h\right) .  \label{34a}
\end{equation}%
Using equation $\left( \ref{34a}\right) $ in $\left( \ref{33a}\right) $ it
takes the form
\begin{equation}
h\left( 1-h\right) \frac{d^{2}X(h)}{dh^{2}}+\left[ c_{2}-\left(
1+a_{2}+b_{2}\right) f\right] \frac{dX(h)}{dh}-a_{2}b_{2}X(h)=0.  \label{35a}
\end{equation}%
In the above we defined
\begin{equation}
a_{2}=\mu _{2}+\nu _{2}+1,  \label{36a}
\end{equation}%
\begin{equation}
b_{2}=\mu _{2}+\nu _{2},  \label{36b}
\end{equation}%
\begin{equation}
c_{2}=1+2\mu _{2}.  \label{36c}
\end{equation}%
Also the constraints coming from the coefficients of $X\left( h\right) $ are
\begin{equation}
\mu _{2}^{2}+\frac{F_{c}^{2}}{B_{c}^{2}}=0  \label{37a}
\end{equation}%
and
\begin{equation}
\nu _{2}^{2}+\frac{F_{c }^{2}}{B_{c}^{2}}-\frac{\lambda _{c}}{B_{c}^{2}}%
=0.  \label{38a}
\end{equation}%
From the above two equations we obtain the values of power coefficients as
\begin{equation}
\mu _{2\pm }=\pm \iota \frac{F_{c}}{B_{c}},\nu _{2\pm }=\pm \sqrt{\frac{F_{c}^{2}}{%
B_{c}^{2}}-\frac{\lambda _{c}}{B_{c}^{2}}}.  \label{39a}
\end{equation}%
Equation $\left( \ref{35a}\right) $ is again a hypergeometric equation and in order to ensure the convergence of the hypergeometric function, we choose negative sign of $\nu_{2}$. Near the cosmological constant both the modes, incoming and outgoing exist, so the general solution can be written as
\begin{equation}
R(h)_{CH}=B_{1}h^{\mu _{2}}(1-h)^{\nu _{2}}F\left(
a_{2},b_{2},c_{2};h\right) +B_{2}h^{-\mu _{2}}(1-h)^{\nu _{2}}F\left(
a_{2}-c_{2}+1,b_{2}-c_{2}+1,2-c_{2};h\right) .  \label{40a}
\end{equation}

\subsection{Matching to an intermediate region}

In order to match the above two solutions of the radial equation i.e.,
near horizon and near cosmological horizon solutions, to an intermediate
region we use matching technique. We first shift the near horizon solution given
in equation $\left( \ref{30}\right) $ to an intermediate region, for which
we change the argument of the hypergeometric function from $f$ to $1-f$ . This
gives the following \cite{SJ, MI}.

\begin{equation*}
R(f)_{NH}=A_{1}f^{\mu _{1}}\left( 1-f\right) ^{\nu _{1}}{\Huge \{}\frac{%
\Gamma (c_{1})\Gamma (c_{1}-a_{1}-b_{1})}{\Gamma (c-a)\Gamma (c-b)}F\left(
a_{1},b_{1},a_{1}+b_{1}-c_{1}+1;1-f\right)
\end{equation*}%
\begin{equation}
+\left( 1-f\right) ^{c_{1}-a_{1}-b_{1}}\frac{\Gamma (c_{1})\Gamma
(a_{1}+b_{1}-c_{1})}{\Gamma (a_{1})\Gamma (b_{1})}F\left(
c_{1}-a_{1},c_{1}-b_{1},c_{1}-a_{1}-b_{1}+1;1-f\right) {\Huge \}}.
\label{41a}
\end{equation}

\bigskip In the limit $r\gg r_{h}$, $f\rightarrow 1$ and we can use
\begin{equation}
\left( 1-f\right) ^{\nu _{1}}\simeq \left( \frac{r_{h}}{r}\right) ^{\nu
_{1}}\simeq \left( \frac{r}{r_{h}}\right) ^{l},  \label{42a}
\end{equation}%
and
\begin{equation}
\left( 1-f\right) ^{\nu _{1}+c_{1}-a_{1}-b_{1}}\simeq \left( \frac{r_{h}}{r}%
\right) ^{2-B_{h}-\nu _{1}}\simeq \left( \frac{r}{r_{h}}\right) ^{-\left(
l+1\right) }.  \label{43a}
\end{equation}%
So, in an intermediate region the solution will take the following form
\begin{equation}
R(r)_{BH}=A_{1}\frac{\Gamma (c_{1})\Gamma (c_{1}-a_{1}-b_{1})}{\Gamma
(c_{1}-a_{1})\Gamma (c_{1}-b_{1})}\left( \frac{r}{r_{h}}\right) ^{l}+A_{1}%
\frac{\Gamma (c_{1})\Gamma (a_{1}+b_{1}-c_{1})}{\Gamma (a_{1})\Gamma (b_{1})}%
\left( \frac{r}{r_{h}}\right) ^{-\left( l+1\right) },  \label{44a}
\end{equation}%
or
\begin{equation}
R(r)_{BH}=\digamma _{1}r^{-\left( l+1\right) }+\digamma _{2}r^{l}.
\label{45a}
\end{equation}%
In the above we have used
\begin{equation}
\digamma _{1}=A_{1}\frac{\Gamma (c_{1})\Gamma (a_{1}+b_{1}-c_{1})}{\Gamma
(a_{1})\Gamma (b_{1})},  \label{46a}
\end{equation}%
\begin{equation}
\digamma _{2}=A_{1}\frac{\Gamma (c_{1})\Gamma (c_{1}-a_{1}-b_{1})}{\Gamma
(c_{1}-a_{1})\Gamma (c_{1}-b_{1})}.  \label{47a}
\end{equation}%
Now we turn to equation $\left( \ref{40a}\right) $ and shift its argument of
hypergeometric function from $h$ to $1-h.$ Thus near cosmological horizon, we have $h(r_{c})\rightarrow 0$, therefore
\begin{equation}
\left( 1-h\right) ^{\nu _{2}}\simeq \left( \frac{r}{r_{c}}\right) ^{\nu
_{2}}\simeq \left( \frac{r}{r_{c}}\right) ^{-\left( l+1\right) },
\label{48a}
\end{equation}%
and
\begin{equation}
\left( 1-h\right) ^{\nu _{2}+c_{2}-a_{2}-b_{2}}\simeq \left( \frac{r}{r_{c}}%
\right) ^{-\left( 1+2\nu _{2}\right) }\simeq \left( \frac{r}{r_{c}}\right)
^{l}.  \label{49a}
\end{equation}%
By using the above approximations and properties of hypergeometric functions
\cite{MI} we can write equation $(\ref{45a})$ as
\begin{equation*}
R(r)_{C}=\left[ B_{1}\frac{\Gamma (c_{2})\Gamma (c_{2}-a_{2}-b_{2})}{\Gamma
(c_{2}-a_{2})\Gamma (c_{2}-b_{2})}+B_{2}\frac{\Gamma (2-c_{2})\Gamma
(c_{2}-a_{2}-b_{2})}{\Gamma (1-a_{2})\Gamma (1-b_{2})}\right] \left( \frac{r%
}{r_{c}}\right) ^{-\left( l+1\right) }
\end{equation*}%
\begin{equation}
\left[ B_{1}\frac{\Gamma (c_{2})\Gamma (a_{2}+b_{2}-c_{2})}{\Gamma
(a_{2})\Gamma (b_{2})}+B_{2}\frac{\Gamma (2-c_{2})\Gamma (a_{2}-b_{2}-c_{2})%
}{\Gamma (a_{2}+1-c_{2})\Gamma (b_{2}+1-c_{2})}\right] \left( \frac{r}{r_{c}}%
\right) ^{l},  \label{50a}
\end{equation}%
or
\begin{equation}
R(r)_{C}=\left( \digamma _{3}B_{1}+\digamma _{4}B_{2}\right) \left( \frac{r%
}{r_{c}}\right) ^{-\left( l+1\right) }+\left( \digamma _{5}B_{1}+\digamma _{6}B_{2}\right) \left( \frac{r}{r_{c}}%
\right) ^{l}.
\label{51a}
\end{equation}%
Here we have used
\begin{equation}
\digamma _{3}=\frac{\Gamma (c_{2})\Gamma (c_{2}-a_{2}-b_{2})}{\Gamma
(c_{2}-a_{2})\Gamma (c_{2}-b_{2})},\digamma _{4}=\frac{\Gamma
(2-c_{2})\Gamma (c_{2}-a_{2}-b_{2})}{\Gamma (1-a_{2})\Gamma (1-b_{2})},
\label{52a}
\end{equation}%
\begin{equation}
\digamma _{5}=\frac{\Gamma (c_{2})\Gamma (a_{2}+b_{2}-c_{2})}{\Gamma
(a_{2})\Gamma (b_{2})},\digamma _{6}=\frac{\Gamma (2-c_{2})\Gamma
(a_{2}-b_{2}-c_{2})}{\Gamma (a_{2}+1-c_{2})\Gamma (b_{2}+1-c_{2})}.
\label{53a}
\end{equation}

\subsection{Greybody factor}

Greybody factor for the emission of scalar fields can be defined by the
amplitudes of the waves at the stretched cosmological horizon solution \cite{PTN}
that is
\begin{equation}
\left\vert A_{l}\right\vert ^{2}=1-\left\vert \frac{B_{2}}{B_{1}}\right\vert
^{2}.  \label{54a}
\end{equation}%
As the power law in $\left( \ref{45a}\right) $ and $\left( \ref{51a}\right) $
is same, we match these two solutions to get the values of $B_{1}$ and $B_{2}
$ as
\begin{equation}
B_{1}=\frac{\digamma _{1}\digamma _{6}-\digamma _{2}\digamma _{4}}{\digamma
_{3}\digamma _{6}-\digamma _{4}\digamma _{5}},  \label{55a}
\end{equation}%
\begin{equation}
B_{2}=\frac{\digamma _{2}\digamma _{3}-\digamma _{1}\digamma _{5}}{\digamma
_{3}\digamma _{6}-\digamma _{4}\digamma _{5}}.  \label{56a}
\end{equation}%
Using values from equations $\left( \ref{55a}\right) $ and $\left( \ref{56a}%
\right) $ in $\left( \ref{54a}\right) $, we get the analytic formula for greybody factor for arbitrary mode$\ l$ as
\begin{equation}
\left\vert A_{l}\right\vert ^{2}=1-\left \vert\frac{\digamma _{2}\digamma _{3}-\digamma
_{1}\digamma _{5}}{\digamma _{1}\digamma _{6}-\digamma _{2}\digamma _{4}}\right\vert ^2.
\label{57a}
\end{equation}

\begin{figure}[H]
  \centering
  \includegraphics[width=0.6\textwidth]{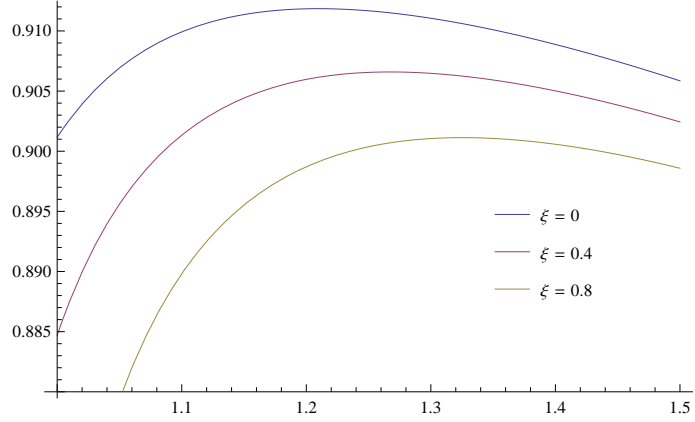}
  \caption{Greybody factor as a function of $\omega r_h$ for different values of $\xi$, when $\Lambda=0.001$, $q=3.5$ and $l=0$.}
  \label{Fig.3a}
  \end{figure}

\begin{figure}[H]
  \centering
  \includegraphics[width=0.6\textwidth]{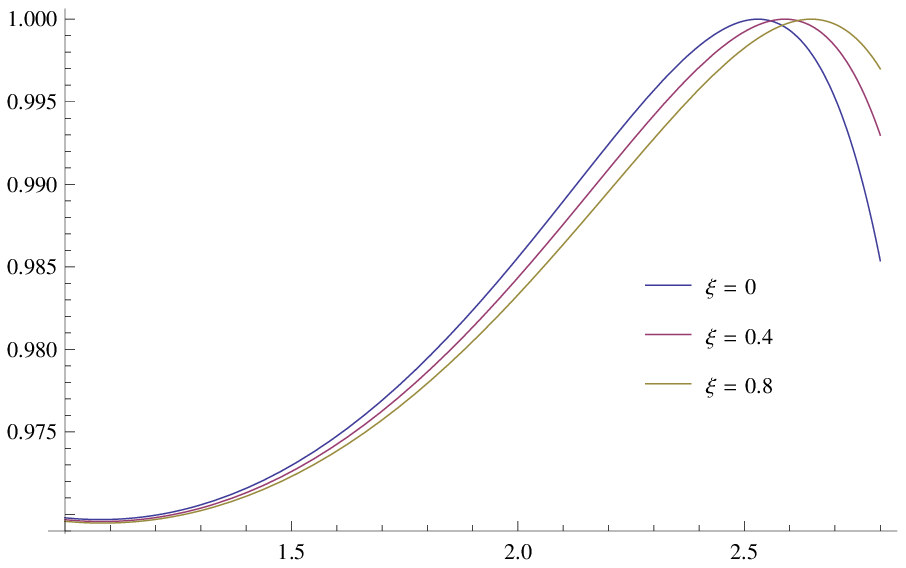}
  \caption{Greybody factor as a function of $\omega r_h$ for different values of $\xi$ , when $\Lambda=0.05$, $q=3.5$ and $l=0$.}
  \label{Fig.3b}
  \end{figure}

\begin{figure}[H]
  \centering
  \includegraphics[width=0.6\textwidth]{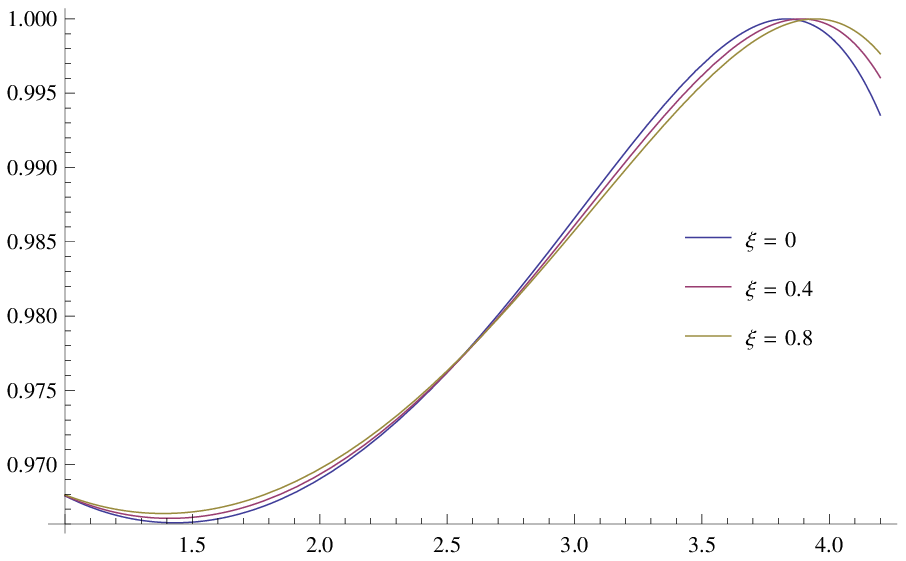}
  \caption{Greybody factor as a function of $\omega r_h$ for different values of $\xi$ , when $\Lambda=0.01$, $q=3.5$ and $l=0$.}
  \label{Fig.3c}
  \end{figure}

\begin{figure}[H]
  \centering
  \includegraphics[width=0.6\textwidth]{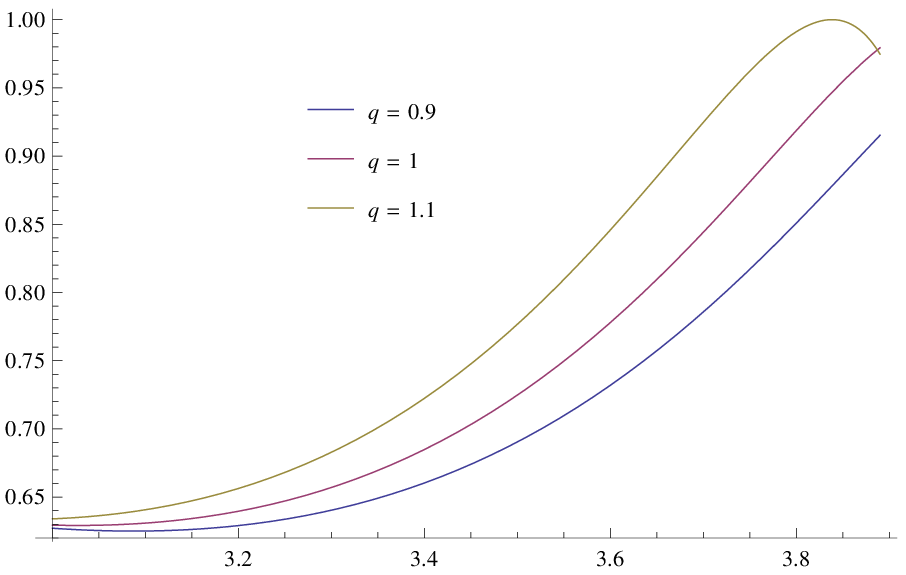}
  \caption{Greybody factor as a function of $\omega r_h$ for different values of $q$, when $\xi=0.2$, $\Lambda=0.001$ and $l=0$.}
  \label{Fig.3d}
  \end{figure}

\begin{figure}[H]
  \centering
  \includegraphics[width=0.6\textwidth]{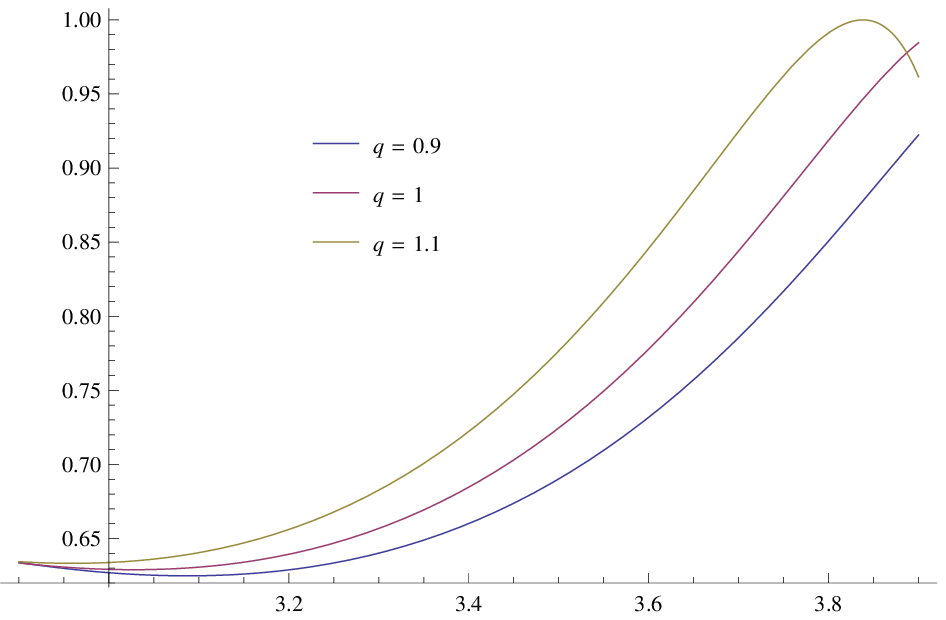}
  \caption{Greybody factor as a function of $\omega r_h$ for different values of $q$, when $\xi=1$, $\Lambda=0.001$ and $l=0$.}
  \label{Fig.3e}
  \end{figure}

In the following plots we depict the greybody factor $\left\vert A_{l}\right\vert ^{2}$ as a function of $\omega r_h$. In Fig. \ref{Fig.3a}, we take $\Lambda=0.001$, $q=3.5$, $l=0$, for different values of $\xi$; in Fig. \ref{Fig.3b} $\Lambda=0.05$, $q=3.5$, $l=0$, for different values of $\xi$; in  Fig. \ref{Fig.3c} $\Lambda=0.01$, $q=3.5$ and $l=0$, for different values of $\xi$; in Fig. \ref{Fig.3d} $\xi=0.2$, $\Lambda=0.001$ and $l=0$, for different values of $q$; in  Fig. \ref{Fig.3e} $\xi=1$, $\Lambda=0.001$, $l=0$, for different values of $q$. An increase in value of the coupling parameter, decreases the greybody factor. This is due to the fact that non-minimal coupling plays the role of an effective mass and hence suppresses the greybody factor.

\subsection{Energy emission}

The flux spectrum, that is, the number of massless scalar particles emitted
by the black hole per unit time is given by \cite{LAE}
\begin{equation}
\frac{dN\left( \omega \right) }{dt}=\frac{d\omega }{2\pi }\frac{1}{e^{\frac{%
\omega }{T_{H}}}-1}\sum_{l=0}^{\infty}\left( 2l+1\right) A_{l}\left(
\omega \right) .  \label{58a}
\end{equation}%
By using the value from equation $\left( \ref{57a}\right) $ into $\left( \ref%
{58a}\right) $ we get
\begin{equation}
\frac{dN\left( \omega \right) }{dt}=\frac{d\omega }{2\pi }\frac{1}{e^{\frac{%
\omega }{T_{H}}}-1}\sum_{l=0}^{\infty}\left( 2l+1\right)\sqrt{ \left( 1-\left\vert\frac{%
\digamma _{2}\digamma _{3}-\digamma _{1}\digamma _{5}}{\digamma _{1}\digamma
_{6}-\digamma _{2}\digamma _{4}}\right\vert ^2\right)} .  \label{59a}
\end{equation}%
Also, the differential energy rate is given by \cite{LAE}
\begin{equation}
\frac{d^{2}E\left( \omega \right) }{dtd\omega }=\frac{1}{2\pi }\frac{\omega
}{e^{\frac{\omega }{T_{H}}}-1}\sum_{l=0}^{\infty}\left( 2l+1\right)
A_{l}\left( \omega \right) .  \label{60a}
\end{equation}%
On using the value of the greybody factor we get
\begin{equation}
\frac{d^{2}E\left( \omega \right) }{dtd\omega }=\frac{1}{2\pi }\frac{\omega
}{e^{\frac{\omega }{T_{H}}}-1}\sum_{l=0}^{\infty}\left( 2l+1\right)\sqrt{\left(
1-\left\vert\frac{\digamma _{2}\digamma _{3}-\digamma _{1}\digamma _{5}}{\digamma
_{1}\digamma _{6}-\digamma _{2}\digamma _{4}}\right\vert ^2\right)}.  \label{61a}
\end{equation}%
The relevance of the low frequency limit is evident from the above results, that
the coupling is only significant in this regime. As the coupling to scalar
field is irrelevant in high frequency limits, the enhancement in
emission rate occurs only at low frequencies.

\subsection{Generalized absorption cross section}

The definition of absorption cross section for asymptotically flat spacetimes is
not valid for asymptotically non-flat spacetimes. For these cases the
general formula for absorption cross section is given by \cite{
LEG,CBF}
\begin{equation}
\sigma =\sum_{l=0}^{\infty}\sigma _{l}=\frac{\pi }{\omega ^{2}}%
\sum_{l=0}^{\infty}\left( 2l+1\right) A_{l}\left( \omega \right) .
\label{62a}
\end{equation}%
Using the value from equation $\left( \ref{57a}\right) $ we get
\begin{equation*}
\sigma =\sum_{l=0}^{\infty}\sigma _{l}=\frac{\pi }{\omega ^{2}}%
\sum_{l=0}^{\infty}\left( 2l+1\right)\sqrt{\left( 1-\left\vert \frac{\digamma _{2}\digamma
_{3}-\digamma _{1}\digamma _{5}}{\digamma _{1}\digamma _{6}-\digamma
_{2}\digamma _{4}}\right\vert ^2\right)}.
\end{equation*}

\section{Conclusion}

In this paper, we have derived the analytic expression of greybody factor for
non-minimally coupled scalar fields from Reissner-Nordstr{\"o}m-de
Sitter black hole in low energy approximation. This expression is valid for
general partial modes. For coupling to scalar curvature, which can be
regarded as mass or charge terms, greybody factor tend to zero like $\omega
^{2}$, irrespective of the values of the coupling parameter. A non-zero greybody factor in low frequency regime means that there is non-zero emission rate of Hawking radiations. The matching technique is used in deriving
the formula for greybody factor. The significance of the results is elaborated
by giving formulae for differential rate of energy and generalized absorption
cross section from greybody factor.

The results of the present study reduce to those of Ref. \cite{PTN} in appropriate
limiting case, that is, if we put charge $Q=0$ we recover the previously
reported results. The effective potential and greybody factor are also analyzed
graphically. We observe that the height of gravitational barrier increases
with the increase of $\xi $, the coupling parameter, whereas in the absence of this parameter it is decreased by increasing values of the cosmological constant. Also from the plot of greybody factor it is observed that an increase in the value of coupling parameter, decreases the greybody factor. This is due to the fact that non-minimal coupling plays the role of an effective mass and hence suppresses the greybody factor. 
Greybody factor is analyzed for different values of charge in the presence of non-minimal coupling.

\section*{Acknowledgement}
KS acknowledges the support provided by the Center for the Fundamental Laws of Nature, Harvard University, during his visit while this work was underway. A research grant from the Higher Education Commission of Pakistan under its Project No. 20-2087 is gratefully acknowledged.

\end{document}